\documentclass[twocolumn,prl,aps,showpacs,floatfix]{revtex4}
\usepackage{graphicx,epsf}

\newcommand{\lsim}{\raise.3ex\hbox{$<$\kern-.75em\lower1ex\hbox{$\sim$}}}
\newcommand{\gsim}{\raise.3ex\hbox{$>$\kern-.75em\lower1ex\hbox{$\sim$}}}

\begin{document}

\title{ Large-Scale Simulations of the Two-Dimensional Melting 
of Hard Disks }

\author{C. H. Mak}

\affiliation{
  Department of Chemistry, 
  University of Southern California, \\
  Los Angeles, California 90089-0482, USA }

\date{\today}

\begin{abstract} 

Large-scale computer simulations involving more than a 
million particles have been performed to study the melting transition in 
a two-dimensional hard disk fluid.
The van der Waals loop previously observed in the pressure-density 
relationship of smaller simulations is shown to be an artifact 
of finite-size effects.  Together with a detailed 
scaling analysis of the bond orientation order, the new results provide 
compelling evidence for the Halperin-Nelson-Young picture.
Scaling analysis of the translational order also yields a 
lower bound for the melting density that is much higher than previously 
thought.

\end{abstract}

\maketitle

A system of hard disks in two dimension (2D) is one of the 
simplest models of a classical fluid.  But beneath the deceptive 
simplicity of this model, 2D hard disks exhibit a set of surprisingly 
rich behaviors.  
Unlike in three dimensions, a 2D solid possesses only 
quasi-long-range translational order which 
decays algebraically to zero at large distances \cite{66mer1133}.
Instead of the usual first-order transition in three dimensions, 
a 2D solid is also expected to melt into a liquid via 
two continuous transitions.
The intervening phase called the ``hexatic'' was predicted 
by Halperin and Nelson \cite{78hal121, 79nel2457} and 
Young \cite{79you1855} to possess quasi-long-range bond orientation 
order but no long-range translational order.  

Given the simplicity of the hard disk model, it would seem easy to 
either prove or disprove the Halperin-Nelson-Young (HNY) theory by 
detailed computer simulation studies.  But twenty-five year after the HNY 
theory was first proposed, simulations that 
could definitively identify the nature of the melting transition 
are still lacking \cite{02bin2323}.  
The first simulation of 2D hard disks
was carried out by Alder and Wainwright \cite{62ald359}.
Based on the appearance of a van der Waals loop in the 
pressure, they concluded that the melting transition must be 
first-order.  Since then, as more computing power has become available, 
simulations have been carried out with increasingly larger system sizes
\cite{68hoo3609, 89zol9518, 92zol11187, 92lee11190, 
95web14636, 97mit6855, 95fer3477, 97fer750, 
98jas277, 99jas134, 99jas2594, 00bat5223, 00sen6294, 02bin2323, 
02mit184202, 02wat041110, 03woj939, 04nie4115, 04jas120, 04wat045103}, 
but instead of clarifying the picture, these simulations have provided 
conflicting conclusions about the nature of melting transition.
One consensus that did emerge from the more recent simulation 
studies is that the 2D hard disk system is very sensitive to finite-size
effects near the melting transition.  This is not unexpected 
if the transition is continuous, but compared to fluids with a soft 
potential \cite{96bag255} the hard disk system is much more prone 
to finite-size errors and boundary effects.
In a simulation of up to $N = 128^2$ particles, Zollweg and 
Chester \cite{92zol11187} observed that the 
equilibration time increased dramatically for densities very close to the
melting transition -- systems of this size were apparently 
not large enough to reach the scaling limit.

\begin{figure}[b]
\includegraphics[width=1.0\columnwidth]{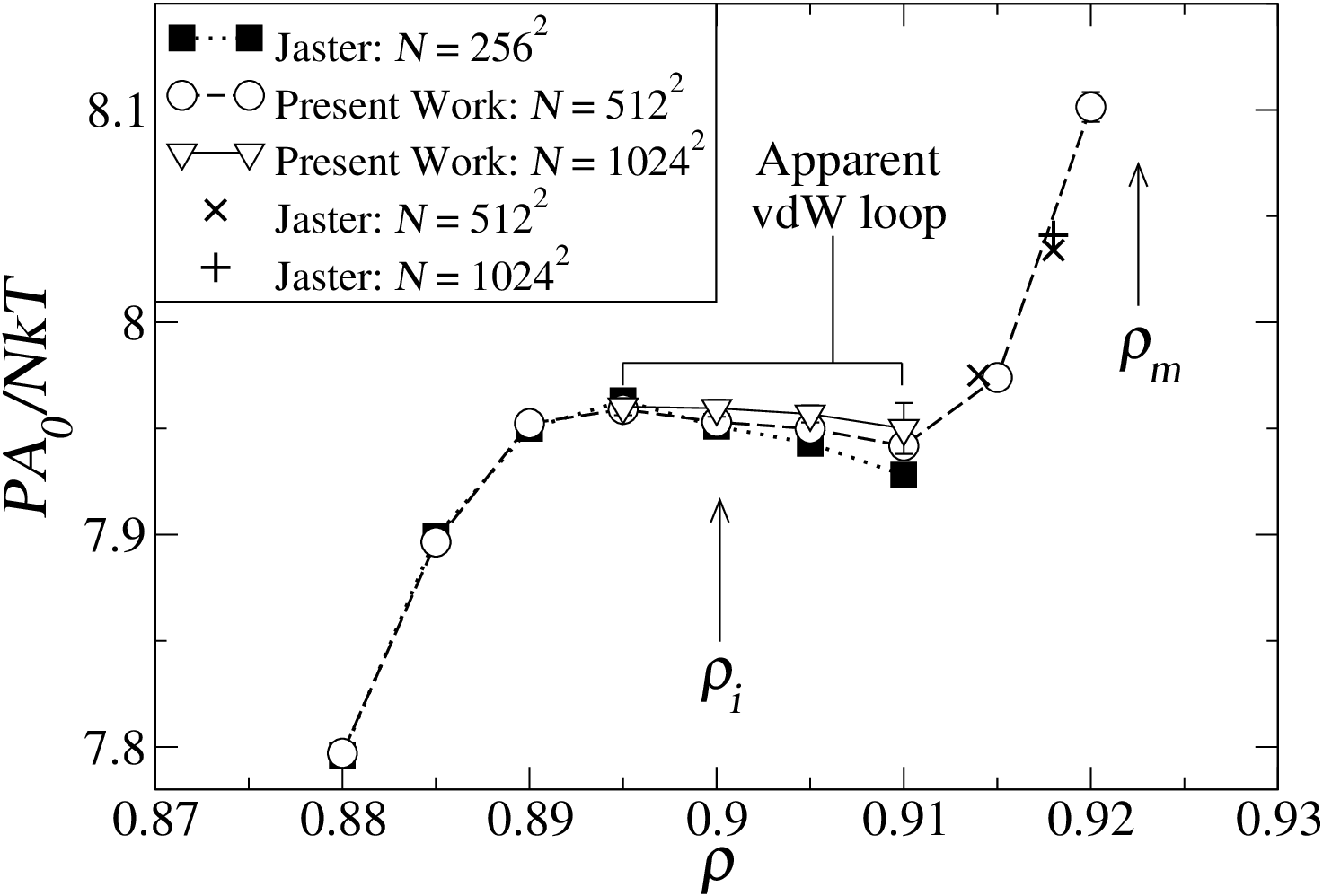}
\caption[]{
Pressure of the hard disk fluid as a function of density.
Solid squares are $N = 256^2$ data from Jaster \cite{99jas2594}, and 
crosses and plus, respectively, are $N = 512^2$ and $N = 1024^2$ 
data from Jaster \cite{04jas120}.
Open circles and open squares are data from the present work for 
$N = 512^2$ and $1024^2$, respectively, with error bars as indicated.
The dotted line is a guide to the eye through Jaster's data for $N = 256^2$.
The dashed and solid line are guides to the eye through the $N = 512^2$ 
and $1024^2$ data in the present work.  
Note the presence of an apparent 
van der Waals (vdW) loop between $\rho$ = 0.895 and 0.910, 
which becomes shallower for increasingly larger size simulations.  
For $N = 1024^2$, the van der Waals loop between $\rho$ = 0.895 and 0.905 has
disappeared completely, with a small decrease in the pressure still visible 
for $\rho$ = 0.910.  
The two arrows indicate the approximate locations of the isotropic-hexatic and 
hexatic-solid boundaries.
}
\end{figure}

The largest simulation that has been performed to date was carried 
out by Jaster with up to $N = 256^2$ particles \cite{99jas134,99jas2594}, 
and more recently for two higher densities with up to 
$N = 1024^2$ \cite{04jas120}.  
Even though a van der Waals loop was observed in the pressure
at densities between $\rho$ = 0.895 and 0.910 (solid squares in Fig.~1), 
Jaster showed using a scaling analysis 
that his data were also compatible with the HNY scenario.
A van der Waals loop is often the sign of a first-order transition, 
but it may also arise from finite-size errors.
To definitively rule out a first-order scenario, one must 
demonstrate that the van der Waals loop is a finite-size artifact, 
i.e. it must be shown to disappear with larger size simulations.
Curiously, the same van der Waals loop was observed for 
two different sizes in Jaster's data -- the pressure for $N = 128^2$ 
(not shown in Fig.~1) and $256^2$ coincide almost completely.

In this letter, we describe a Monte Carlo study of 2D hard disks for 
up to $N = 1024^2 = 1048576$ particles.  
The calculations were carried out in the
canonical ensemble, in a square box with periodic boundary 
condition and in a rectangular box with aspect ratio $\sqrt{3}:2$ 
for the higher densities.
We worked with densities in the range $\rho = 0.880$ to $0.920$, which 
according to previous estimates should span the transition region
\cite{62ald359, 92zol11187, 95web14636, 99jas2594, 02bin2323}.  
Densities $\rho$ are given in reduced units where the hard disk diameter 
is one.

While the rationale for going to larger system size is to 
eliminate finite-size effects, larger simulations also take longer 
to equilibrate.  
We have focused 
on $N = 512^2$ to try to carry out detailed simulations covering a large 
range of densities between $\rho$ = 0.880 and 0.920.
At this size, one run at each density took several months of CPU time.
Additional larger simulations with $N = 1024^2$ were performed for 
four densities between $\rho$ = 0.895 and 0.910 in the vicinity of 
the van der Waals loop previously observed in smaller simulations.
In contrast, Jaster's recent simulations \cite{04jas120} focuses on 
a different region in the phase diagram, 
offering data for $\rho$ = 0.918 at $N = 1024^2$ 
and two densities, $\rho$ = 0.914 and 0.918, for $N = 512^2$.  

Two different types of Monte Carlo moves were used for our simulations.  
The first is a conventional Metropolis move, where each particle is displaced
in a random direction by a random amount.  
A second Monte Carlo move based on the cluster
algorithm proposed by Dress and Krauth \cite{95dre597} and 
Liu and Luiijten \cite{04liu035504} 
was also used.  
At the densities we worked with, neither algorithm is 
particularly efficient in causing very large rearrangements in the system
configuration.  But by mixing two different algorithms that have vastly 
different properties, we hope to minimize equilibration problems 
characteristic of any single algorithm.  One Monte Carlo step (MCS) in 
our simulation is defined as having moved each particle on the average 
once using the Metropolis algorithm, plus having made one global cluster 
update.  
The simulations reported here were carried out with no fewer than 
5 million MCS for each density.  Depending on the equilibration rate, 
results from the last 1 to 3 million MCS were used to collect statistics.
Two to four independent simulations were carried out for each density for 
simulations with a square box, and five to six for those with a 
rectangular box.  

The pressure $P$ was calculated using the virial formula $PA_0/NkT =
[1 + \pi\rho g(1^+)/2] \sqrt{3}\rho /2$, where $g(1^+)$ is the 
contact value of the pair correlation function and $A_0 = \sqrt{3}N/2$ 
is the closed-packed area of the system.  
The calculated pressure $P$ is shown in Fig.~1 as a function of 
density $\rho$ for $N = 512^2$ (open circles) and for $N = 1024^2$ 
(open triangles).  
Comparing the $N = 512^2$ and $1024^2$ data to
those from Jaster's simulation with $N = 256^2$, the two sets of data 
are almost identical for $\rho \leq 0.890$, but inside the range $\rho$ 
= 0.895 to 0.910, the larger size simulations produced 
a smaller pressure for $\rho$ = 0.895 but larger 
pressures for $\rho$ = 0.900 to 0.910. 
It is therefore clear that the apparent van der Waals loop in the pressure 
is a result of finite-size effects, and using even larger size simulations, 
this slight nonmonotonic decrease in the pressure should eventually vanish 
altogether.
For the $N = 1024^2$ simulations, the van der Waals loops has 
completely disappeared between $\rho$ = 0.895 and 0.905, with a slight 
dip in $P$ still visible for $\rho$ = 0.910.  
As expected, finite-size 
effects are indeed very pronounced in the transition region even for 
simulations of this magnitude.  

Even though the evidence in Fig.~1 is compelling that the 
van der Waals loop is an artifact of finite-size effects, these data 
alone cannot definitively rule out a first-order melting transition.
For this, wee need to carefully analyze the finite size effects.
To disentangle the finite-size effects, a 
detailed scaling analysis must be performed on the simulation data.  
We have found a subblock scaling analysis 
\cite{95web14636, 96bag255} to be useful for this purpose.  
With this method, a single large size simulation provides information on 
multiple length scales simultaneously.  
The subblock scaling analysis was applied to the bond orientation 
order as well as the translational order.

The bond orientation order is given by 
$\psi_6^2 = \vert (6N)^{-1} \sum_l \sum_j \exp(6i\theta_{lj})\vert^2$, 
where the sum goes over each particles $l$ and its 
nearest neighbors $j$ and $\theta_{lj}$ is the angle between
the line from $l$ to $j$ with some fixed reference axis.  
According to HNY theory, $\psi_6$ 
should have only short-range order in the isotropic phase and
quasi-long-range order in the hexatic phase with exponent $\eta_6 \leq 1/4$. 
We calculated $\psi_6$ for subblock sizes of $L_B = L/64$, $L/32$ 
$\ldots$ $L$, where $L$ is the full length of the box and plot the 
results in Fig.~2.  For $\rho \leq 0.895$, $\psi_6$ clearly 
scales to zero, but for $\rho \geq 0.900$, $\psi_6$ appears to scale to a 
finite value.

\begin{figure}
\includegraphics[width=1.0\columnwidth]{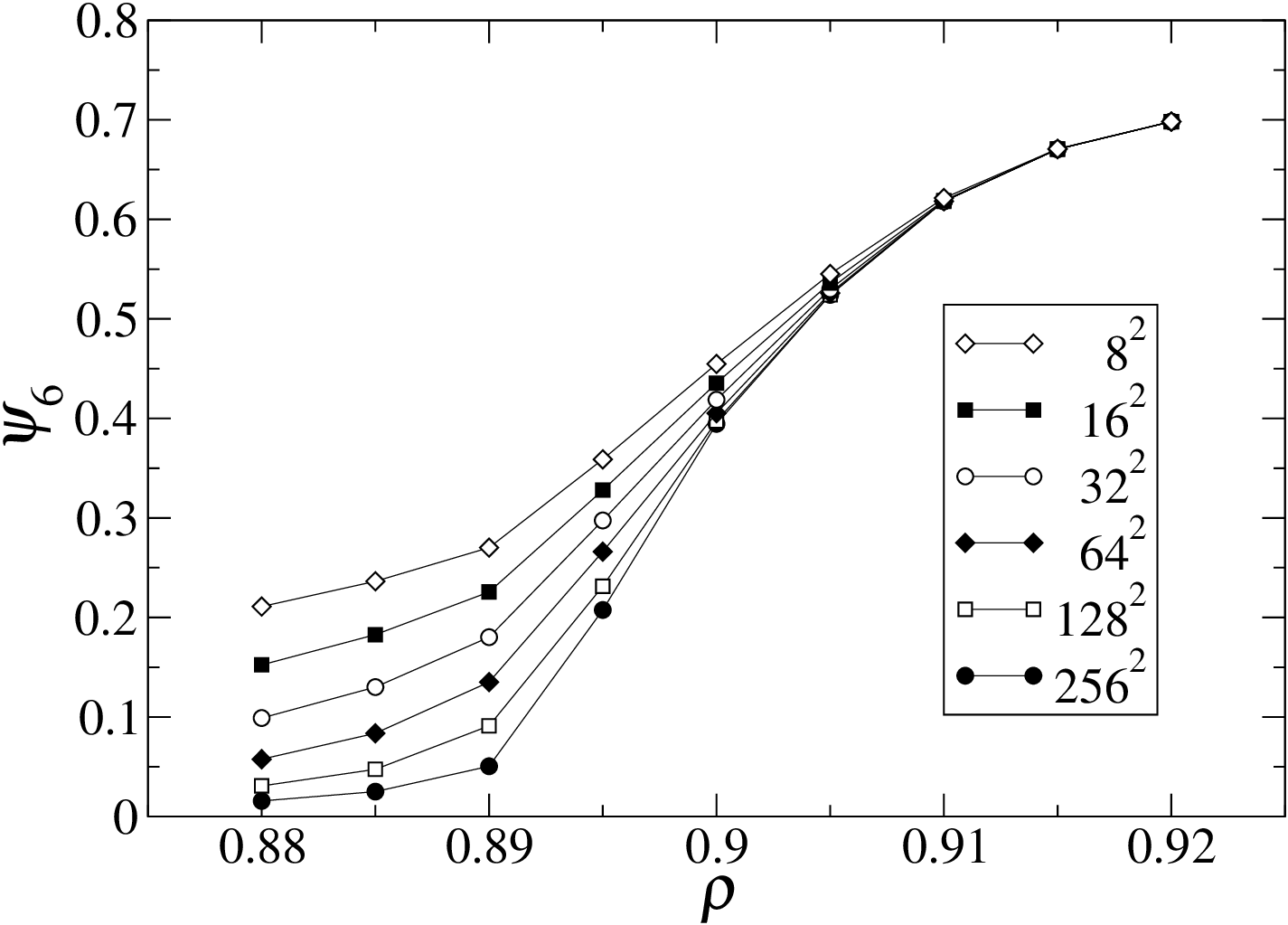}
\caption[]{
The bond orientation order parameter $\psi_6$ derived from the subblock
analysis as a function of density for the $N = 512^2$ simulations.  
For $\rho \leq 0.895$, $\psi_6$ scales 
to 0 with larger system sizes.  For $\rho \geq 0.905$, $\psi_6$ appears 
to scale to a nonzero value.
}
\end{figure}

To establish the precise scaling behavior, we plot 
$\ln \psi_6^2$ vs. the natural log of the length of the subblock $L_B$ 
in Fig.~3 for the $N = 512^2$ simulations.
According to HNY theory, this plot should show a slope $-\eta_6$ in
the hexatic phase and $-2$ in the isotropic phase where there is only 
short-range order.
Figure~3 shows that for both $\rho$ = 0.880 (solid triangles) and 0.890 
(open diamonds), the bond 
orientation order has no long-ranged correlations in the long 
length scale limit, and the size of the simulations was large enough to
reach the scaling limit.
We can safely conclude that 
densities $\rho \leq 0.890$ are in the isotropic phase.
On the other hand, for the highest densities $\rho$ = 0.905 
(solid diamonds) and 0.910 (open squares), 
the bond orientation order shows an algebraic decay with an exponent 
$\eta_6$ much smaller than 1/4.  This is consistent with the 
interpretation that these densities are either inside the hexatic or the
solid phase.

\begin{figure}
\includegraphics[width=1.0\columnwidth]{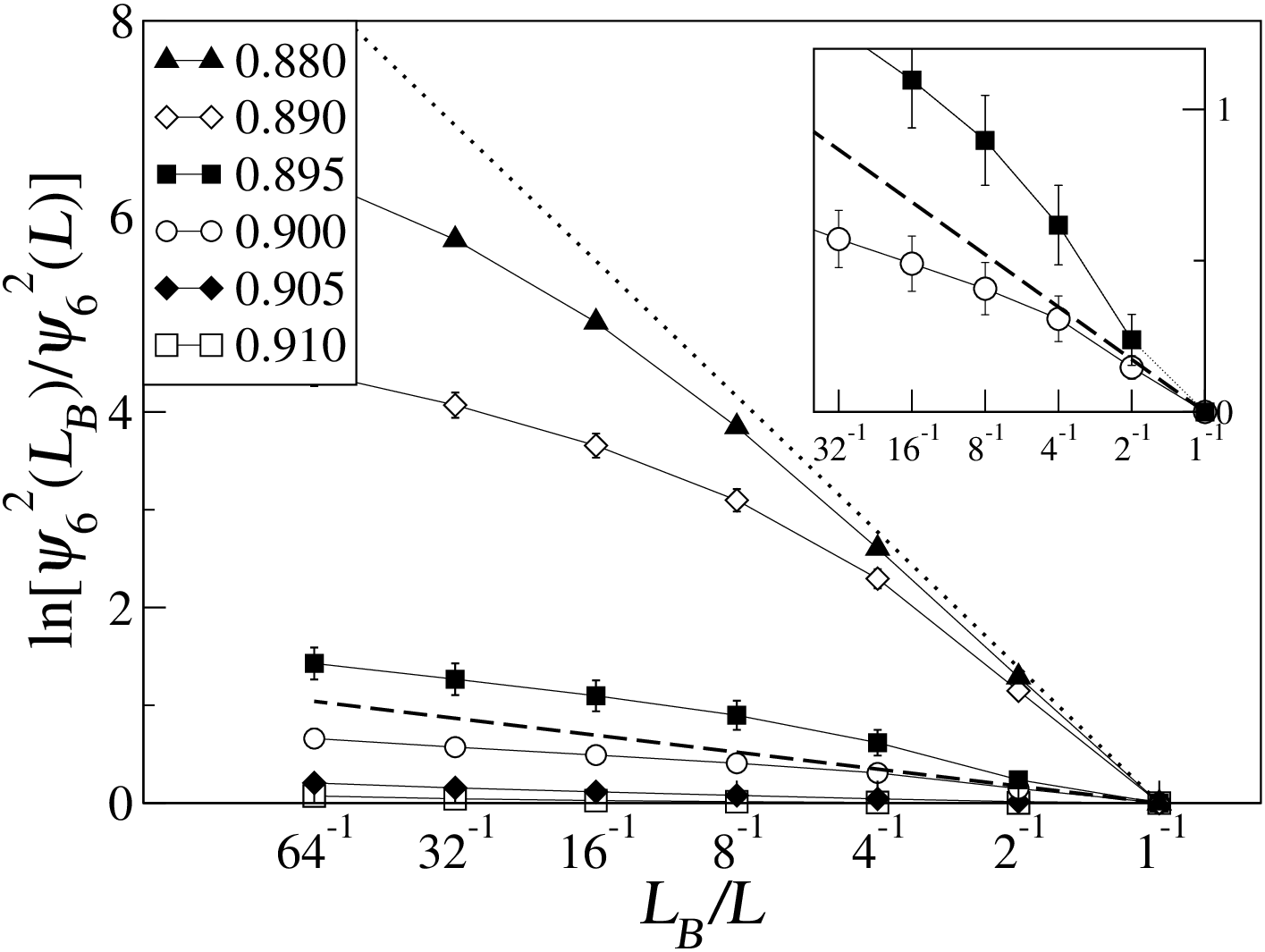}
\caption[]{
Subblock scaling analysis for the bond orientation order parameter 
for the $N = 512^2$ simulations.  
The dotted line corresponds to a slope of $-2$ and the dashed line 
a slope of $-1/4$.  The inset shows an expanded view 
for $\rho$ = 0.895, 0.900 and 0.905 in the large length scale region.
}
\end{figure}

For the two densities $\rho$ = 0.895 and 0.900, the interpretation of the 
subblock scaling plots is more involved.  
The inset in Fig.~3 shows an expanded 
view of their behaviors in the large length scale limit.
For $\rho$ = 0.900 (open circles), the bond orientation
order shows a slope that is very close to $-1/4$ at large length scales.  
In the HNY scenario, this is consistent with a density inside the hexatic 
phase, very close to the hexatic-isotropic boundary $\rho_i$.
These evidence suggest that $\rho_i \lsim 0.900$.

For $\rho$ = 0.895 (closed squares), the subblock scaling plot 
changes slope twice, first at $L/2$ and then more gradually 
between $L/4$ and $L/8$.
The first abrupt slope change at $L/2$ is 
an artifact of the subblock scaling analysis which has been discussed 
by Weber, Marx and Binder \cite{95web14636}.  
The reason for this sudden slope change 
is that the subblocks and the full box actually 
belongs to two different ensembles -- the 
canonical for the full box and something resembling the grand 
canonical for the subblocks.
It is therefore 
possible for the full box to exhibit a different scaling behavior 
compared to the subblocks when the correlation length exceeds the 
size of the simulation box, in which case the full box data point must be 
excluded from the scaling analysis.  When this is done, the scaling behavior 
suggests that the orientation order decays 
algebraically with an exponent larger than $1/4$.  But clearly the 
scaling limit has not been reached, so it is possible that this
exponent will continue to increase with lengths beyond the size of 
the present simulation.
These evidence suggest that $\rho$ = 0.895 must still be inside the 
isotropic phase but is very close to the isotropic-hexatic boundary.

Taken together, the pressure data and the subblock scaling 
analysis of the bond orientation order reveal a consistent picture.  
For densities $\rho \leq 0.895$, the
system is in the isotropic phase.  
The van der Waals loop in the pressure between $\rho$ = 0.895 and 0.910 
observed in previous simulations is most certainly due to finite-size effects.
The bond orientation 
correlation length increases when 
the isotropic-hexatic boundary $\rho_i$ is approached from below 
and it changes 
from short-range correlation to an algebraic decay with 
$\eta_6$ close to 1/4 at $\rho_i \lsim 0.900$, 
which is consistent with previous estimates \cite{99jas2594}.  
Above $\rho_i$, the exponent $\eta_6$ decreases quickly from 1/4 
to zero when the hexatic-solid boundary $\rho_m$ is approached from below.
These findings are consistent with the HNY scenario.

The fact that $\eta_6 \to 0$ for $\rho \to 0.910$ has been used 
previously to suggest that the hexatic-solid boundary is 
at $\rho_m \approx 0.910$ \cite{95web14636,99jas2594}.  
The recent data of Jaster, however,  
have placed $\rho_m$ at a much higher value $\approx$ 0.933 \cite{04jas120}.
To more accurately locate the hexatic-solid boundary $\rho_m$, we turn to 
a subblock scaling analysis of the translational order 
$\psi_t^2 = \vert N^{-1} \sum_l \exp(i \vec{k}\cdot\vec{r}_l)\vert^2$,
where the wavevector $\vec{k}$ has magnitude $2\pi/(\sqrt{3}/2\rho)^{1/2}$.  
In the solid phase, $\psi_t^2$ is expected to decay algebraically with 
exponent $\eta_t$ = 1/3.
The results for three densities, $\rho$ = 0.900, 0.910 and 0.920, are 
shown in Fig.~4 for the $N = 512^2$ simulations in both a square and a 
rectangular box with a $\sqrt{3}:2$ aspect ratio.
For $\rho$ = 0.900 (triangles) and 0.910 (squares), 
the results are consistent with no long-range 
translational order in the large length scale limit for both box 
geometries.  This indicates that both of these densities are 
inside the hexatic phase.  
On the other hand, for $\rho$ = 0.920, 
the translational order shows apparently 
different scaling behaviors for the two box geometries -- no long-range 
order in the square box but quasi-long-range 
translational order with an apparent exponent $\eta_t > 1/3$ for the 
rectangular box.  In fact, 
comparing the two different box geometries, 
we found that the rectangular box simulations at this density 
were much slower to equilibrate,  
leading to the larger error bars on the right panel of Fig.~4 
for $\rho$ = 0.920.
Since the translational correlation length is expected 
to diverge according to HNY theory as $\rho$ approaches the melting 
density $\rho_m$ \cite{79nel2457}, 
the rectangular box used for the simulations at $\rho$ = 0.920 was 
probably too small to reach the scaling limit.  
Therefore, we believe that $\rho = 0.920$ is most likely still 
inside the hexatic phase and has not yet reached the hexatic-solid boundary.  
This establishes a lower bound for $\rho_m$,  
one that is much higher than the value previously 
suggested \cite{95web14636,99jas2594}.  But this new 
lower bound is consistent with the recent estimate provided by Jaster based 
on simulations with $N$ up to $1024^2$ \cite{04jas120}.
Since the pressure at $\rho$ = 0.920 (see Fig.~1) is much higher than
the pressure inside the apparent van der Waals loop, 
this new lower bound for $\rho_m$ also 
provides evidence corroborating the conclusion we have 
drawn from the size-dependence of the pressure-density data in Fig.~1 
that the apparent van der Waals loop in the pressure is not related to a 
first-order transition.  

\begin{figure}
\includegraphics[width=1.0\columnwidth]{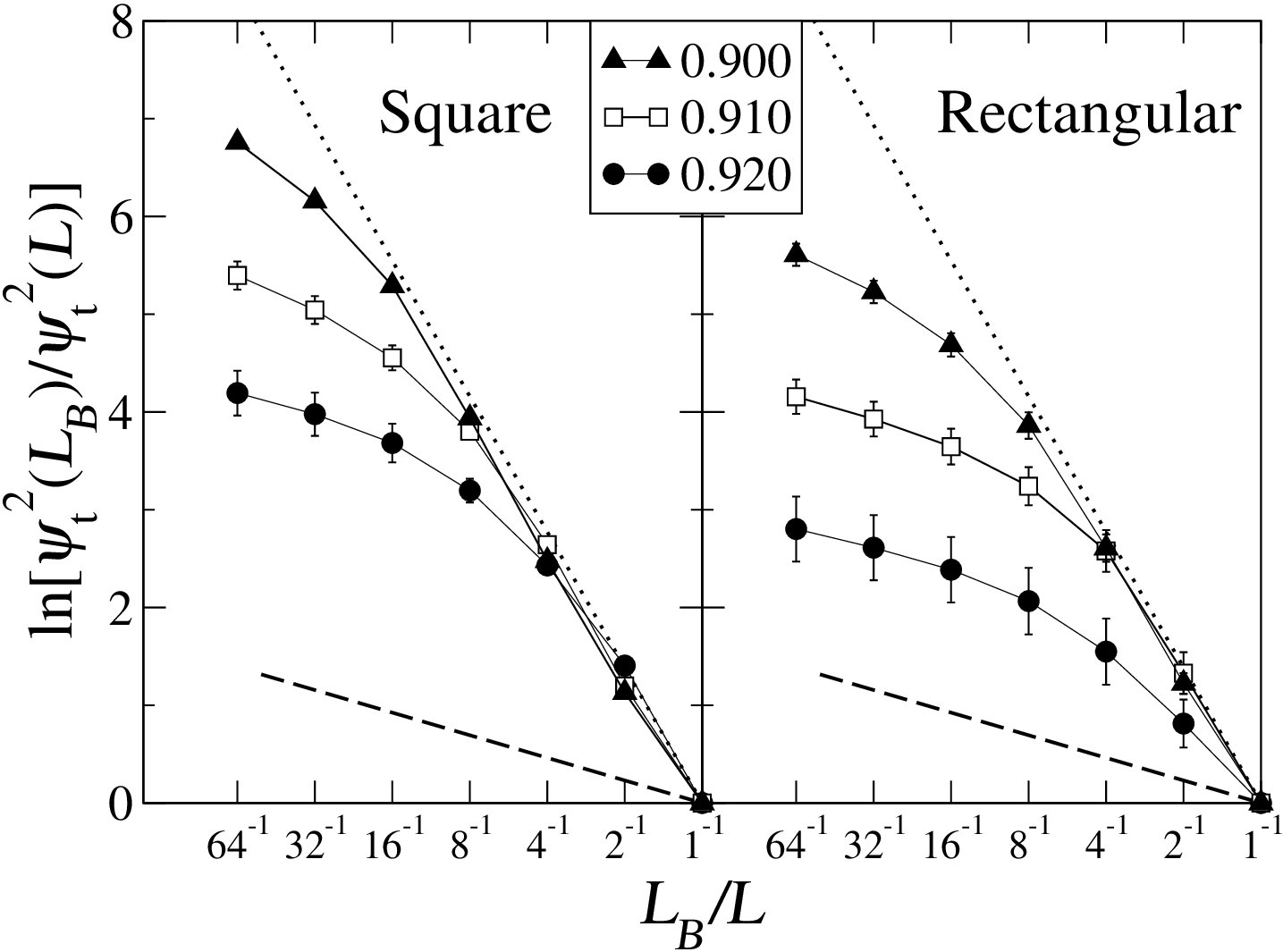}
\caption[]{
Subblock scaling analysis for the translational order parameter for 
the $N = 512^2$ simulations in a square box and a rectangular box.
The dotted line corresponds to a slope of $-2$ and the dashed line 
a slope of $-1/3$.
}
\end{figure}

In conclusion, we have shown using 
large-scale computer simulations with more than a 
million particles that the apparent van der Waals loop observed 
previously in smaller simulations is an artifact of finite-size effects. 
In conjunction with a detailed scaling analysis, the data provide 
compelling evidence for a continuous isotropic-hexatic transition 
as predicted by HNY theory at $\rho_i \lsim 0.900$.  
Scaling analysis of the translational order also yields a 
lower bound for the melting density, $\rho_m > 0.920$, one that is much higher 
than previously thought, providing additional evidence that the apparent 
van der Waals loop is not due to a first-order transition.

\begin{acknowledgments}

This work was supported by the National Science Foundation under grant 
CHE-9970766.  The author has benefited from helpful discussions with 
Hans C. Andersen.

\end{acknowledgments}

\end{document}